# Deterministic multivalued logic scheme for information processing and routing in the brain

Sergey M. Bezrukov [a] and Laszlo B. Kish [b,1]

[a] *Laboratory of Physical and Structural Biology, Program in Physical Biology, NICHD, National Institutes of Health, Bethesda, MD 20892, USA*

[b] *Department of Electrical and Computer Engineering, Texas A&M University, Mailstop 3128, College Station, TX 77843-3128, USA*

**Abstract.** Driven by analogies with state vectors of quantum informatics and noise-based logic, we propose a general scheme and elements of neural circuitry for processing and addressing information in the brain. Specifically, we consider random (e.g., Poissonian) trains of finite-duration spikes, and, using the idealized concepts of excitatory and inhibitory synapses, offer a procedure for generating $2^N - 1$ orthogonal vectors out of $N$ partially overlapping trains ("neuro-bits"). We then show that these vectors can be used to construct $2^{2^N - 1} - 1$ different superpositions which represent the same number of logic values when carrying or routing information. In quantum informatics the above numbers are the same, however, the present logic scheme is more advantageous because it is deterministic in the sense that the presence of a vector in the spike train is detected by an appropriate coincidence circuit. For this reason it does not require time averaging or repeated measurements of the kind used in standard cross-correlation analysis or in quantum computing.

---

[1] Until 1999: L.B. Kiss

# 1. Introduction

Alexander Graham Bell patented the telephone invention on March 10, 1876 [1]. The first telephones were rented in pairs which could only talk with each other because they were connected by direct individual lines. However, only two years later, in 1878, one of the first commercial switchboards serving 21 telephones began operating in Boston, Massachusetts. From this moment in history, telephone signals started to be routed through a reduced number of lines thus drastically decreasing the costs of building and maintenance.

In the human body neuronal signals may also travel relatively long distances before they get to the right place in the nervous system [2]. How are they routed? Different parts of the brain have multiple connections with each other and with the periphery. Quite clearly there is a significant hierarchy which makes neuronal network drastically different from the one of the first telephone networks mentioned above. However, does our nervous system use something analogous to the 1878 switchboard?

With the hundred billion neurons in the brain acting in a massively parallel mode, routing of information is the crucial part of its processing. The design of neuronal networks displays complex topological features used to achieve the necessary connectivity patters [3] whose knowledge allows researchers to approach the long-standing problems of information encoding [4, 5].

Here we offer a hypothetical scheme that allows routing and, simultaneously, provides a multivalued deterministic logic framework for information processing and logic operations in the brain. Briefly, our idea is that the information and/or the addresses can be encoded in a superposition of orthogonal random trains or "basis vectors" of nerve spikes. We show that using idealized elements of nerve circuitry such as excitatory and inhibitory synapses, it is possible to generate $2^N - 1$ basis vectors out of *N* initial partially overlapping random (e.g., Poissonian) trains. The transduction cable then consists of one line used for the given binary (on/off) superposition of the basis vectors plus *N* parallel lines for the initial trains. It is easy to see that such a scheme reduces the total number of lines in comparison with the direct connection via individual lines by a factor of $(2^N - 1)/(N + 1)$. For example, using 7 initial trains it is possible to simultaneously communicate with 127 "customers" through 8 lines only, thus saving the number of parallel lines by more than an order of magnitude.

To detect the presence of a particular vector in the superposition, the receiving logic unit generates $2^N - 1$ vectors identical to those in the transducing part and uses them in a set of coincidence circuits to analyze the actual superposition and decide about the further processing or addressing. The coincidence circuits operate deterministically in the sense that they do not require averaging of the kind used in cross-correlation analysis of neuronal signals [2] or repeated measurements, averaging, and extracting statistics inherent to quantum informatics. In this respect our scheme is deterministic. Clearly, communicating binary messages to $2^N - 1$ parallel different addresses, is equivalent to $2^{2^N - 1} - 1$ logic values.

**2. Multidimensional logic state vectors from neural spike trains: The framework**

In this section, we show the mathematical framework for constructing such a logic system operating on *N* partially overlapping Poisson-like neural spike trains. First we briefly describe the noise-based logic hyperspace with similar properties, which is based on independent stochastic processes with zero mean. Then we show how these goals can be realized with unipolar neural spikes.

**2.1 The noise-based logic hyperspace.** Very recently, a system based on the so-called *logic hyperspace* vectors of noise-based logic was introduced [6], in which the logic values are constructed from independent electronic noises represented by stochastic processes with zero mean. Having $N$ independent noise processes $U_i(t)$ ($N$ *noise-bits*, where $1 \leq i \leq N$) [7]), and supposing binary "on/off" (0 or 1) choices for each noise-bit, it is possible to construct $2^N - 1$ different products of these noises. For example, with $N = 3$, one has the following $7$ ($= 2^3 - 1$) different products:

$$U_1(t), U_2(t), U_3(t), U_1(t)U_2(t), U_1(t)U_3(t), U_2(t)U_3(t), \text{ and } U_1(t)U_2(t)U_3(t). \quad (1)$$

These products are all orthogonal to each other [6, 7], where orthogonality means that the product of any two of these products has zero mean value. Therefore, the $2^N - 1$ different products represent orthogonal basis vectors of the $2^N - 1$ dimensional logic space instead of the original space with $N$ dimensions. Using the bracket notation of quantum informatics, the products can be represented by binary (bit) vectors, where at the location of the "on" bits we put 1 and at the location of the "off" bits we put 0. Thus the above system of products will become:

$$|1,0,0\rangle, |0,1,0\rangle, |0,0,1\rangle, |1,1,0\rangle, |1,0,1\rangle, |0,1,1\rangle, |1,1,1\rangle, \quad (2)$$

respectively. Here symbol 0 means multiplication by 1. Using these orthogonal state vectors linear superpositions can be constructed [6, 7]:

$$S = \sum_{i=1}^{2^N-1} a_i \left| c_{1,i}, c_{2,i}, \ldots, c_{N,i} \right\rangle, \quad (3)$$

where the $a_i$ ($i = 1,\ldots,2^N$) and the $c_{m,n}$ ($m = 1,\ldots,N$; $n = 1,\ldots,2^N$) binary coefficients can have either 0 or 1 values. A given superposition $S$ represents a distinct logic value and, in the present binary case, we have

$$M = 2^{2^N-1} - 1 \quad (4)$$

different logic values. For $N = 3$, $M = 127$ which should be compared to 8 if regular bits would have been used instead of noise-bits. According to Eq. (4) the number of logic values grows fast with the dimensionality of the original space, and for $N = 5$ it gives about *2.1 billion* different logic values which should be compared to 32 provided by regular bits.

These properties are very similar to the Hilbert space representation of quantum informatics, however the quantum physical collapse of wavefunction does not exist here, thus it does not hinder the performance [7]. As an illustration, a string search algorithm for a $2^N$ arbitrary physical string has been proposed where the noise-based engine finds the string during a single clock period [7].

## 2.2 Orthogonal logic space from partially overlapping neural spike trains.

The situation with neural spike trains is significantly different, and it needs a different approach. The mean value of a spike train is non-zero and, therefore, the products of two different uncorrelated neural spike trains is non-zero either. This means that they are not orthogonal. Furthermore, analog multipliers used to produce the orthogonal set in list (1) do not exist.

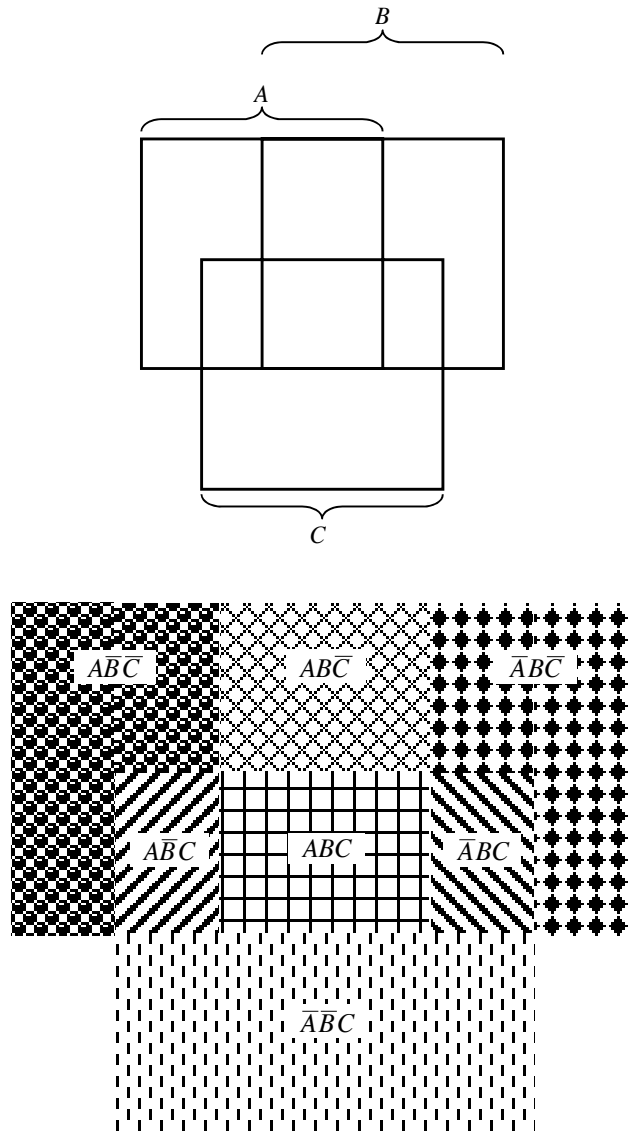

**Figure 1.** Schematic illustration of the seven orthogonal trains prepared from three overlapping spike trains (neuro-bits) *A, B, and C*, respectively. The overlapping spike trains are represented by three overlapping squares (upper figure).

To build orthogonal logic vectors from partially overlapping spike trains, or *neuro-bits*, we must use a set-theoretical approach. In the upper part of Figure 1, partially overlapping squares *A*, *B*, and *C* represent the system of three partially overlapping neural spike trains. In the lower part of Figure 1 we show 7 non-overlapping (orthogonal) subsets that cover the whole system. The 7 subsets represent orthogonal vectors of a 7 dimensional logic space. Though neuro-bits A, B, C are not orthogonal, the hyperspace obtained by this set-theoretical approach has a 7 dimensional orthogonal vector basis.

Using the bracket representation of quantum informatics we can specify the orthogonal logic vector basis as follows. The system of orthogonal states (non-overlapping sub-sets) corresponding to lists (1) and (2) above is naturally:

$$\begin{aligned}|1,0,0\rangle &= A\overline{B}\overline{C}, \\ |0,1,0\rangle &= \overline{A}B\overline{C} \\ |0,0,1\rangle &= \overline{A}\overline{B}C \\ |1,1,0\rangle &= AB\overline{C} \\ |1,0,1\rangle &= A\overline{B}C \\ |0,1,1\rangle &= \overline{A}BC \\ |1,1,1\rangle &= ABC\end{aligned} \quad , \tag{5}$$

where the neuro-bits can have 0 or 1 values. The remaining question is if the set theory approach, which requires generation of intersections (products) and complementary sets, can be implemented with idealized elements of the neuronal circuitry. In the next section, we introduce the basic circuits of the deterministic brain logic making this task achievable.

## 3. Neuron circuitry to generate and utilize the logic state vectors and superpositions

The following circuits, built out of idealized single neurons and idealized synapses, are the simplest logic gates. They serve as elements of more complex logic circuits, and their potential applicability is not limited to the illustrative examples given below.

We call *N-th order orthogonator gate* the neuron circuitry that has $N$ input ports for $N$ partially overlapping neural spike trains, neuro-bits, and $2^N - 1$ output ports providing $2^N - 1$ orthogonal (non-overlapping) sub-sequences of the input spikes. The circuit utilizes both the excitatory and the inhibitory inputs of the neuron [2]. If the excitatory and inhibitory inputs receive spike trains $A(t)$ and $B(t)$, respectively, then the output will provide $A(t)\overline{B}(t)$ where the product and upper bar mean set-theoretical intersection and complement operations. Thus $A(t)\overline{B}(t)$ contains those elements of $A(t)$ which do not overlap with $B(t)$.

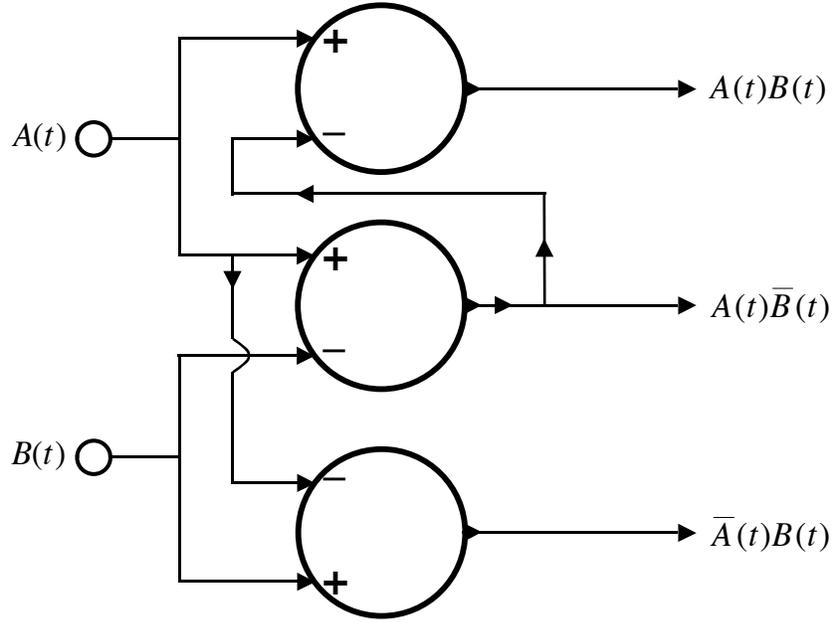

**Figure 2.** The second-order *orthogonator* gate circuitry utilizing both excitatory (+) and inhibitory (-) synapses of neurons. The input points at the left are symbolized by circles and the output points at the right by free-ending arrows. The arrows in the lines show the direction of signal propagation.

In Figure 2, a 2-nd order orthogonator gate circuit is shown. The spike trains $A(t)$ and $B(t)$ with partial overlap provide three orthogonal outputs: $A(t)\overline{B}(t)$, $\overline{A}(t)B(t)$, $A(t)B(t)$. With these two neuro-bits, the number of orthogonal outputs is $2^2 - 1 = 3$ and the number of different logic values these orthogonal vectors can form in a binary superposition is $2^{2^2-1} - 1 = 7$.

To discuss higher order orthgonators and other neural circuitry, here we introduce the *orthon* gate, which is a part of the circuit in Figure 2, see Figures 3, 4. The orthon can be used in any brain circuit where multiplication (set-theoretical intersection) $AB$ is needed and it can also provide another output with $A\overline{B}$ where $A$ and $B$ are the neural spike trains feeding its (+) and (-) inputs, respectively.

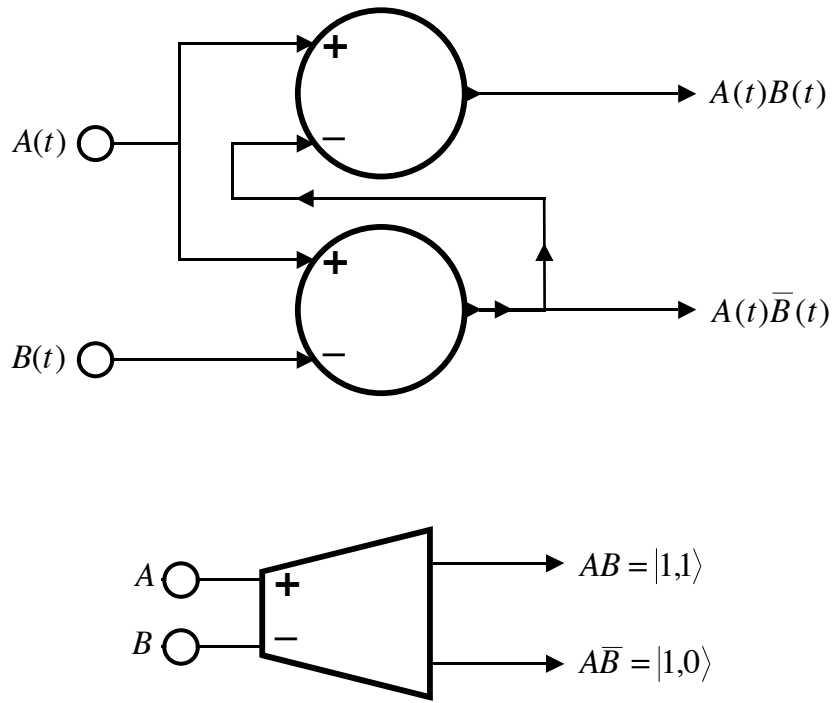

**Figure 3.** The *orthon* gate, neural building element (top) and its symbol (bottom). Arbitrary orders of *orthgonators* can be built with orthons. However, for the simplest design they must be combined with single neurons.

In Figure 4, the 2-nd order orthogonator is shown by the use of an orthon and a neuron. It is the same circuit as shown in Figure 2 where the orthon contains the upper two neurons.

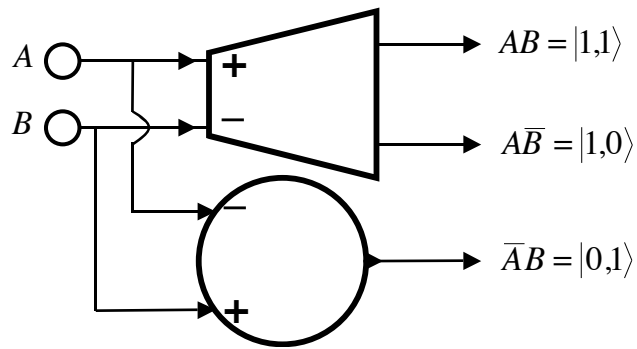

**Figure 4.** The 2-nd order *orthogonator* consists of 1 *orthon* and 1 neuron.

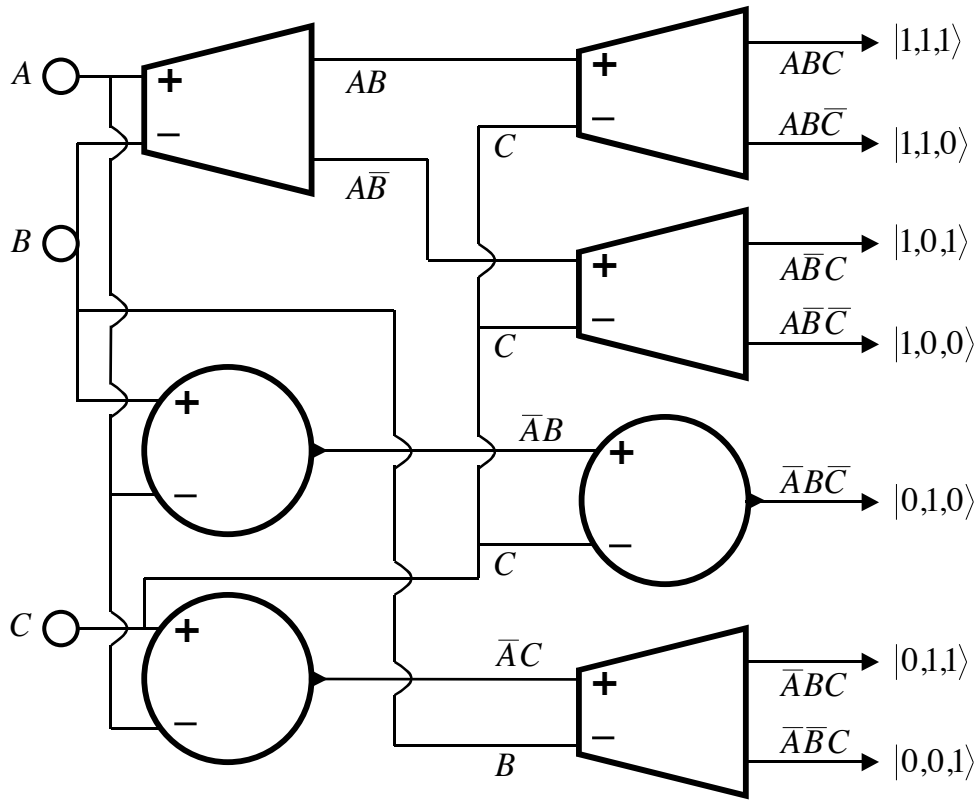

**Figure 5.** The third order orthogonator consists of 4 orthons and 3 neurons, that is, altogether 11 neurons.

The 3-rd order orthogonator gate, which is the neural circuit implementation of the set-theoretical operation illustrated in Figure 1, is shown in Figure 5. At the left, the three neuro-bit inputs are driven by partially overlapping neural spike trains while the 7 outputs on the right give orthogonal, non-overlapping sub-sequences shown also with the bracket representation of quantum informatics. This leads to the 127 different logic values (superpositions) mentioned above.

The circuits for addition are well known [2]. It could be a single neuron with several parallel excitatory inputs. Because the input is supplied by orthogonal (non-overlapping) spike trains, the output will be the superposition of the input sequences.

In Figure 6, another use of the orthon gate is shown, the one in which it plays a role of a projection operator, similar to the projection operator of quantum electrodynamics. It can detect an orthogonal basis vector $R_k = |c_{1,k}, c_{2,k}, \ldots, c_{N,k}\rangle$ in a superposition of Eq. (3), and projects or restores it at its upper output. The lower output will provide the superposition without that component. It is important to point it out that to detect the vector in the superposition, *no time averaging or another way of collecting statistics is needed*. In the circuitry built of idealized elements considered here, the presence of vector $R_k = |c_{1,k}, c_{2,k}, \ldots, c_{N,k}\rangle$ in the superposition is detected as soon as the lower input receives the first spike of this vector.

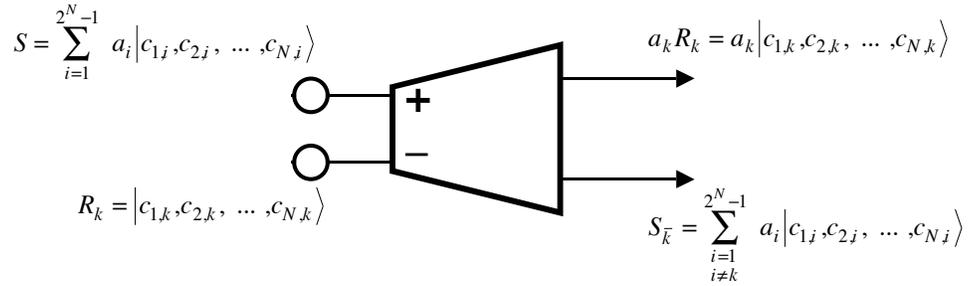

**Figure 6.** The projection operation circuitry for addressing and associative memory tasks (similarly to string search in noise-based logic) needs a single orthon. The upper output extracts the *k*-th basis vector from the superposition, indeed if it exists in the superposition. No averaging is necessary, it is a coincidence type of detection; the first output spike there reports on the presence of the basis vector in the superposition. The lower output will not contain the *k*-th vector but only the rest of the input superposition.

As a final example, Figure 7 shows a "spectrum analyzer" circuit built of orthons, which detects all the existing basis vectors $R_k = |c_{1,k}, c_{2,k}, \ldots, c_{N,k}\rangle$ in the input superposition $S = \sum_{i=1}^{2^N - 1} a_i R_i$. The circuit has a similar role as quantum Fourier transformation in quantum informatics [8], and, for this reason, may be called *neural Fourier transformation*.

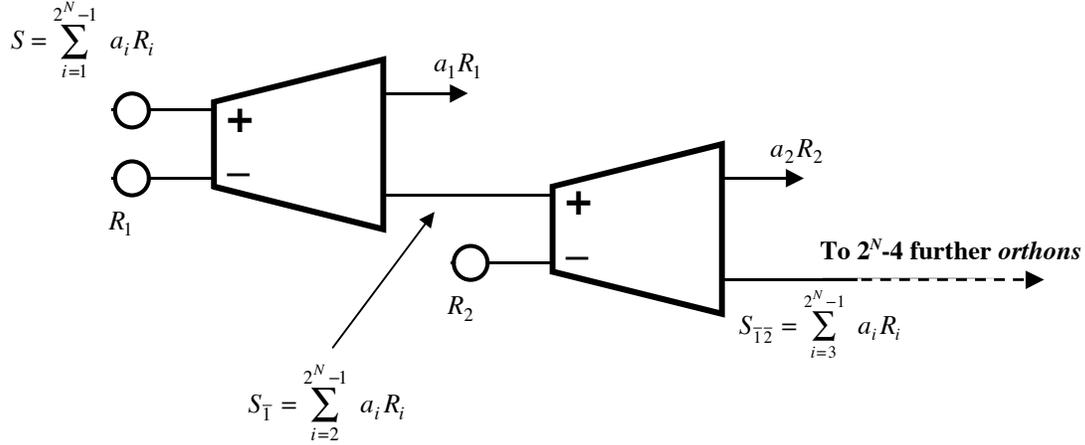

**Figure 7.** "Neural Fourier transformation" analogous to quantum Fourier transformation. $2^N - 2$ orthons used as projection operators to get the full spectrum of orthogonal base vectors in the input superposition. The last orthon's lower output will provide the detection of the last superposition vector. Note, the number of necessary orthons can be radically decreased if only specific superpositions are needed to be detected.

In an alternative realization, the (+) inputs of the orthons can be jointly driven by the analyzed superposition. The disadvantage of the solution in Figure 7 is accumulation of time delay; however its advantage is that each lower output reduces the whole superposition by one basis vector therefore the required number of orthons is less by 1. Then the last orthon's lower output will do the display/detection of the last sought superposition vector.

Concerning the completeness of this logic scheme it is relevant to ask here the following question of opposite nature. Instead of the presence of an orthogonal basis vector, how fast can the present idealized logic detect the lack of a given base component in a superposition? The answer is straightforward: as soon as in the relevant reference vector the first spike arrives and the logic realizes that there is no coinciding spike in the superposition. That means that an inverting operation is as fast as the relevant regular logic function.

The completeness of noise-based logic was proven in the introductory paper [6] by defining its universal AND, OR and INVERTER gates. Because our present brain logic scheme is also based on an orthogonal vector base, in its idealized form it is also complete. The practical aspects of the logic including completeness, error rate and error propagation will be studied in forthcoming works.

Our goal here was not to explore all the possibilities offered by the scheme of the orthogonal multidimensional logic space. Rather, it was to suggest a potential role of the given examples in building brain processors with multivalued logic functions and gates for decision making, addressing, routing, and receiving addressed information. The examples are just a few basic ones to demonstrate their feasibility. The basic Boolean logic functions for the gates with noise-based logic are given in [6] and the very same functions also work here, whenever binary logic is needed in brain functioning.

## 4. Conclusions

We find a rich *multivalued*, *deterministic* logic scheme which can be realized using a few random, partially overlapping neural spike trains and does not require the heavy statistical analysis involved in schemes of general purpose quantum informatics [9]. Certainly, the randomness of initial neural spike trains leads to random delays in the action of coincidence circuits. However, the delays are small enough to be compatible with the speed of brain functioning. For example, assuming that the maximal number of spikes per second in a single axon is of the order of one hundred [2] and that only one axon is used for the superposition transduction, the case of generating 7 basis vectors out of 3 initial spike trains considered in Section 2.2, will provide an average delay of about 100 ms.

The logic scheme described in our study provides an efficient way to route and encode information. Importantly, it is based on existing, though idealized, elements and uses random pulse sequences for the generation of orthogonal vectors. Indeed, if one would construct hypothetical encoding algorithms without the restrictions known to exist in the brain, such as narrow bandwidth and stochastic timing, much higher information transfer rates could be achieved. For example, if one allowed for the existence of an infinitely wide bandwidth and an exact master clock with infinitely precise timing in pulse generation and reading in the brain, the stationary information transfer rate could be made arbitrary high in a similar fashion as in pulse-position modulation (PPM) used in optical communication [10].

Moreover, even for random Poisson-like sequences, a scheme where a master clock combined with a sufficiently fast switchboard separates neuronal pulse sequences into shorter subsequences and routes them between proper addresses would be equal in efficiency to the logic scheme described above. However, the absence of such a master clock is recognized as one of the major features of the brain that makes it so different from a digital computer [11].

To "understand… mathematical operations employed by neural circuits" [11] is one of the imperatives of modern neuroscience. We hope that the proposed scheme, inspired by analogies with quantum informatics and noise-based logic, can be of help in advancing in this direction.

**Acknowledgment**

SMB is indebted to Adrian Parsegian for fruitful suggestions and reading parts of the manuscript; LBK appreciates discussions with Sunil Khatri, Mark Dykman, Ferdinand Peper, Swaminathan Sethuraman and Suhail Zubairy. This study was supported by the Intramural Research Program of the Eunice Kennedy Shriver National Institute of Child Health and Human Development, National Institutes of Health. SMB's visit of Texas A&M University (January 2009) and LBK's visit of NIH (November 2008) was partly supported by a TAMU subcontract of the Army Research Office grant under contract W911NF-08-C-0031. LBK's visit was also supported by a TAMU travel grant for sabbatical leave.